# Control Flow Duplication for Columnar Arrays in a Dynamic Compiler

Sebastian Kloibhofer[a], Lukas Makor[a], David Leopoldseder[b], Daniele Bonetta[c], Lukas Stadler[b], and Hanspeter Mössenböck[a]

a Institute for System Software, Johannes Kepler University Linz, Austria

b Oracle Labs, Austria

c Oracle Labs, Netherlands

**Abstract** Columnar databases are an established way to speed up online analytical processing (OLAP) queries. Nowadays, data processing (e.g., storage, visualization, and analytics) is often performed at the programming language level, hence it is desirable to also adopt columnar data structures for common language runtimes.

While there are frameworks, libraries, and APIs to enable columnar data stores in programming languages, their integration into applications typically requires developer interference. In prior work, researchers implemented an approach for *automated* transformation of arrays into columnar arrays in the GraalVM JavaScript runtime. However, this approach suffers from performance issues on smaller workloads as well as on more complex nested data structures. We find that the key to optimizing accesses to columnar arrays is to identify queries and apply specific optimizations to them.

In this paper, we describe novel compiler optimizations in the GraalVM Compiler that optimize queries on columnar arrays. At JIT compile time, we identify loops that access potentially columnar arrays and duplicate them in order to specifically optimize accesses to columnar arrays. Additionally, we describe a new approach for creating columnar arrays from arrays consisting of complex objects by performing *multi-level storage transformation*. We demonstrate our approach via an implementation for JavaScript `Date` objects.

Our work shows that automatic transformation of arrays to columnar storage is feasible even for small workloads and that more complex arrays of objects could benefit from a multi-level transformation. Furthermore, we show how we can optimize methods that handle arrays in different states by the use of duplication. We evaluated our work on microbenchmarks and established data analytics workloads (TPC-H) to demonstrate that it significantly outperforms previous efforts, with speedups of up to 10x for particular queries. Queries additionally benefit from multi-level transformation, reaching speedups of around 2x. Additionally, we show that we do not cause significant overhead on workloads not suitable for storage transformation.

We argue that automatically created columnar arrays could aid developers in data-centric applications as an alternative approach to using dedicated APIs on manually created columnar arrays. Via automatic detection and optimization of queries on potentially columnar arrays, we can improve performance of data processing and further enable its use in common—particularly dynamic—programming languages.

## 1 Introduction

Database management systems have long proven the significance of optimized data layouts to improve query performance on large, complex data [1, 2, 3, 11, 32, 47, 49, 79, 83]. Columnar databases [4, 13, 48, 78] have emerged as the de-facto standard for read-heavy query execution, e.g., in OLAP (online analytical processing), where queries target large, multi-dimensional, and typically read-only data [21]. In a columnar database, the records of a table are decomposed to store all values per column in a contiguous memory region. This improves the performance of queries that target a limited amount of columns as the column values of individual records may be efficiently processed in tight loops, leading to improved cache and memory efficiency.

Nowadays, data-intensive operations are often conducted on the language level, with runtimes and compilers taking on similar roles as database query optimizers [7, 14, 35, 38, 54, 57, 63, 75]. However, while there are specific APIs that provide columnar storages in programming languages [7, 56, 75], these shift the responsibility of choosing the right memory layout for the right case to the application developers.

Based on these observations, Makor et al. [55] developed an approach for *automated* transformation of arrays of objects to columnar storage in the GraalVM JavaScript runtime [66]. They detect suitable arrays at run time and transform the underlying storage to a columnar layout if the array is large enough and frequently accessed. Custom compiler optimizations in the GraalVM Compiler [29, 53, 77] allow them to achieve performance benefits after just-in-time (JIT) compilation.

Consider the function `shippingCosts` in Figure 1a that takes an array of `clients` to calculate the applicable shipping costs per client. Figure 1b depicts the memory layout of `clients`, with the individual objects scattered throughout the heap. In the aforementioned approach, the array is transformed to columnar storage at some point during the execution, resulting in the memory layout shown in Figure 1d. The function is subsequently optimized during compilation. Figure 1c depicts a high-level representation of the compiled function. Now, a run-time check—a so-called *guard*—ensures that the array is indeed columnar, such that execution of the compiled code is aborted if this is not the case (cf. Section 2.1). This enables additional optimizations that remove now redundant memory accesses and move operations out of the loop.

One problem of this approach is that the compiler optimizations rely on the assumption that they can focus on a single array state, e.g., that the optimized accesses always target columnar arrays. A method is therefore either compiled for columnar or for non-columnar arrays. Calling a compiled method with an array in an unexpected state triggers *deoptimization* [41] (cf. the guard in the example), i.e., execution falls back to the interpreter and the compiled code is discarded. However, this process causes additional overhead due to recompilation that impacts smaller, less frequently executed workloads. Furthermore, this behavior is undesirable in JIT-compiled environments, where ideally compilation of *hot* methods would stabilize after a while.

Additionally, workloads using arrays of objects with reference type properties don't perform well, as accesses to these properties represent most of the run-time effort and cannot be optimized as much. This problem particularly manifests itself when dates and strings are used in queries.

**(a)** A function that calculates shipping costs for an array of clients based on their location.

```
function shippingCosts(clients) {
  const cost = Array(clients.length).fill(0);
  for (let i = 0; i < clients.length; i++)
    cost[i] = clients[i].zip == 4040 ? 15 : 20;
  return cost;
}
```

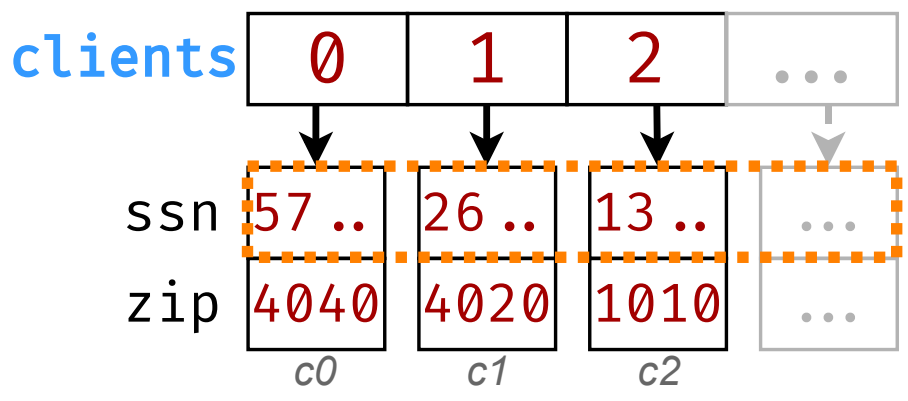

**(b)** Memory representation of the array before transformation.

**(c)** High-level representation of the function after columnar storage transformation on clients.

```
function shippingCosts(clients) {
  const cost = Array(clients.length).fill(0);
  guard <isColumnar>(clients);
  const zip = clients._propArrays["zip"];
  for (let i = 0; i < clients.length; i++)
    cost[i] = zip[i] == 4040 ? 15 : 20;
  return cost;
}
```

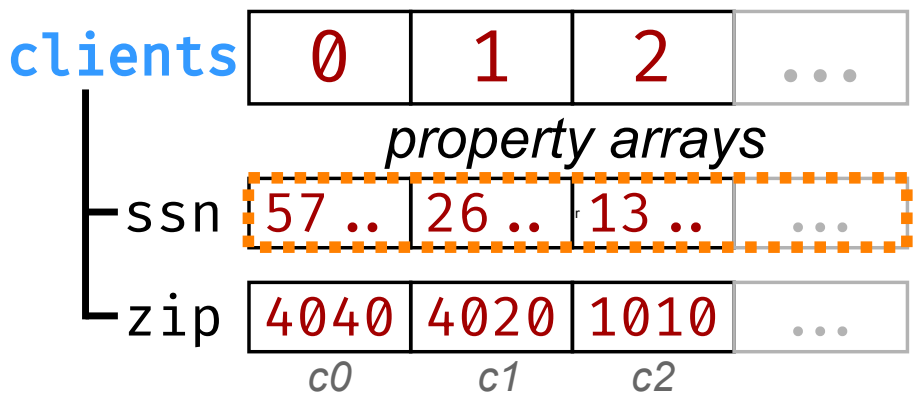

**(d)** Memory representation of the array after transformation.

**Figure 1** Transformation of an array of clients.

Therefore, we propose a novel approach that detects optimizable queries at JIT compile time and subsequently uses code duplication to enable optimization of loops over columnar arrays. Additionally, we demonstrate a new approach to handle nested data within columnar arrays via a *multi-level storage transformation*.

Overall, this work contributes the following:

1. Novel compiler optimizations for the approach proposed in [55] that automatically duplicate the control flow for accessing columnar arrays to mitigate recompilation efforts and to speed up smaller workloads (Section 4).
2. A refinement of the storage transformation approach that improves the performance of accessing array-of-object data structures containing objects with date properties; this approach may be expanded to other complex nested types (Section 3).
3. An evaluation of our refinements based on microbenchmarks and TPC-H queries [23] that shows performance improvements of up to 3x for smaller workloads and up to 10x for larger workloads (Section 5).

Additionally, Section 2 provides necessary background about the compiler and the approach this work is based on. Section 6 describes limitations of our approach. In Section 7, we discuss related work.

## 2 Background

GraalVM [65] is a high-performance Java virtual machine. The Truffle language implementation framework [39, 85, 88]—a GraalVM component for implementing abstract syntax tree (AST) interpreters—enables execution of other programming languages. We integrated our proposed approach into the Truffle-based, standard-compliant JavaScript runtime GraalVM JavaScript [66] and the GraalVM Compiler.

### 2.1 GraalVM Compiler

The GraalVM Compiler [29, 53, 77] is a self-hosting, high-performance dynamic JIT compiler that optimizes programs using run-time profiling information. The compiler applies speculative optimizations based on run-time observations to make assumptions about a method (e.g., about types, values) and to enable follow-up optimizations [30, 31]. If an assumptions fails, the runtime uses *deoptimization*. Deoptimization allows the runtime to abort execution of a compiled method, discard the compiled code, and resume execution in the interpreter [41].

**Graal IR** In the GraalVM Compiler, methods are first transformed into an intermediate representation known as Graal IR [29]. Graal IR is a structured, directed graph consisting of nodes that represent the control flow and the data flow of a method. The IR is in *static single assignment form* [19, 25, 59, 60, 84], with $\phi$ nodes representing values that depend on the current path (e.g., a $\phi$ node captures an assignment whose value is branch-dependent, a loop $\phi$ represents values modified in a loop). To enable traversal of the IR during compilation, Graal IR nodes are typically scheduled within *basic blocks* [20, 70]. Basic blocks are a minimal, branchless set of nodes and are connected to their corresponding predecessors and successors.

The so-called *dominator relation* [6, 22, 25, 37, 51, 74] imposes a partial order upon the basic blocks of a method: A basic block $B_0$ *dominates* another basic block $B_1$ if every path on the control flow to $B_1$ traverses $B_0$ first. Using the *direct dominator* of each basic block (i.e., the closest dominating block), the basic blocks form the *dominator tree*, where the root block represents the method entry. Hence, compiler phases frequently utilize this structure for traversing the nodes of a method.

During compilation, the IR is modified in a variety of compiler phases. These include partial escape analysis [77], inlining [73], and SIMD vectorization [15, 24, 43, 44]. Finally, the compiler generates highly-optimized machine code from the IR.

### 2.2 Truffle

Truffle [39, 85, 88] is a language implementation framework that enables guest-language integration via AST interpretation. At run time, the guest-language ASTs are partially evaluated and optimized based on profiling information accumulated during interpretation [42, 50, 87, 89] and are subsequently compiled using the GraalVM Compiler. Truffle furthermore provides a standardized object storage model [86] that can be used across all language implementations.

Our approach is based on the high-performance JavaScript Truffle implementation GraalVM JavaScript [66]. GraalVM JavaScript is ECMAScript [33] compliant and features a Node.js backend [69]. GraalVM JavaScript makes use of the Truffle object storage model [86] to support the JavaScript object and array semantics [33].

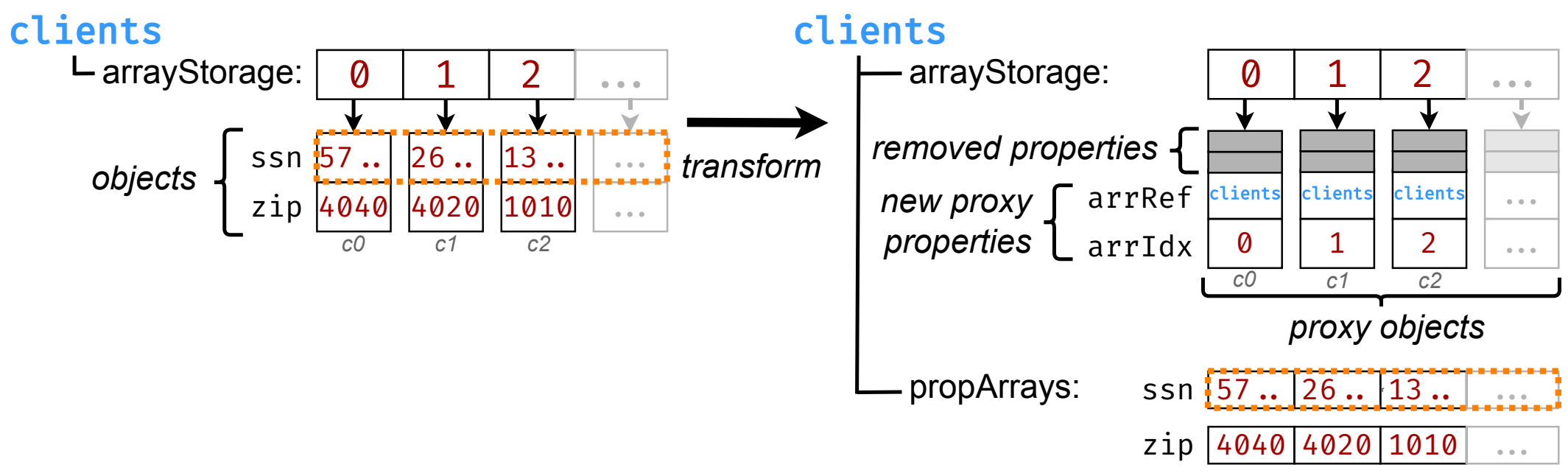

Figure 2 Transforming an array of objects to a columnar array (based on [55]).

## 2.3 Columnar Arrays in GraalVM JavaScript

Makor et al. [55] showed that data-intensive queries on analytical workloads can be significantly accelerated by transforming arrays of uniform objects to columnar arrays and optimizing accesses to the resulting columnar storage accordingly. Such a representation typically improves cache-utilization when iterating over the array and furthermore allows optimization of array accesses in loops at compile time as shown in [55]. To perform such a transformation, information about arrays of objects is tracked at run time. When a fitting and frequently used array that only contains objects of the same type is detected (cf. clients on the left-hand side of Figure 2), the transformation to a columnar array is initiated. First, a verification step ensures that the array indeed only holds viable objects (all objects have the same type, their properties are accessible, etc.). Otherwise, the transformation step is aborted. If successful, one new array is created for each property of the objects in the original array. Then, all the values of the objects for the respective property are copied to the new array. Hence, these arrays are called *property arrays*. As depicted at the bottom right of Figure 2, the original array subsequently references these property arrays.

Each object in the original array is transformed to a *proxy object*. As stated in [55], the name *proxy* is adopted from related work [56, 71] and is not to be confused with ECMAScript proxy objects [33]. The proxy transformation is necessary, because after the array transformation, each object has to access the columnar storage when it needs to access its data. To access the property arrays that make up the columnar storage, the proxy object needs to access the array to load the correct property array and requires the index of the array element to access the property array at the correct position. Hence, every proxy object contains the original array reference (arrRef) and the index of the object in the array (arrIdx), as shown in the top right of Figure 2. The transformation to a proxy object is only expressed via an internal type descriptor change that reflects the object's new role, hence object identity and type checks are not affected and JavaScript object semantics are preserved. If unsupported operations are performed on a columnar array, a restoration is performed, i.e., the array is transformed back to the non-columnar state. Such unsupported operations include inserting incompatible elements, introducing holes, but also adding or deleting properties of the array elements or assigning new values that are incompatible with the corresponding property array (e.g., assigning a string to a numeric property).

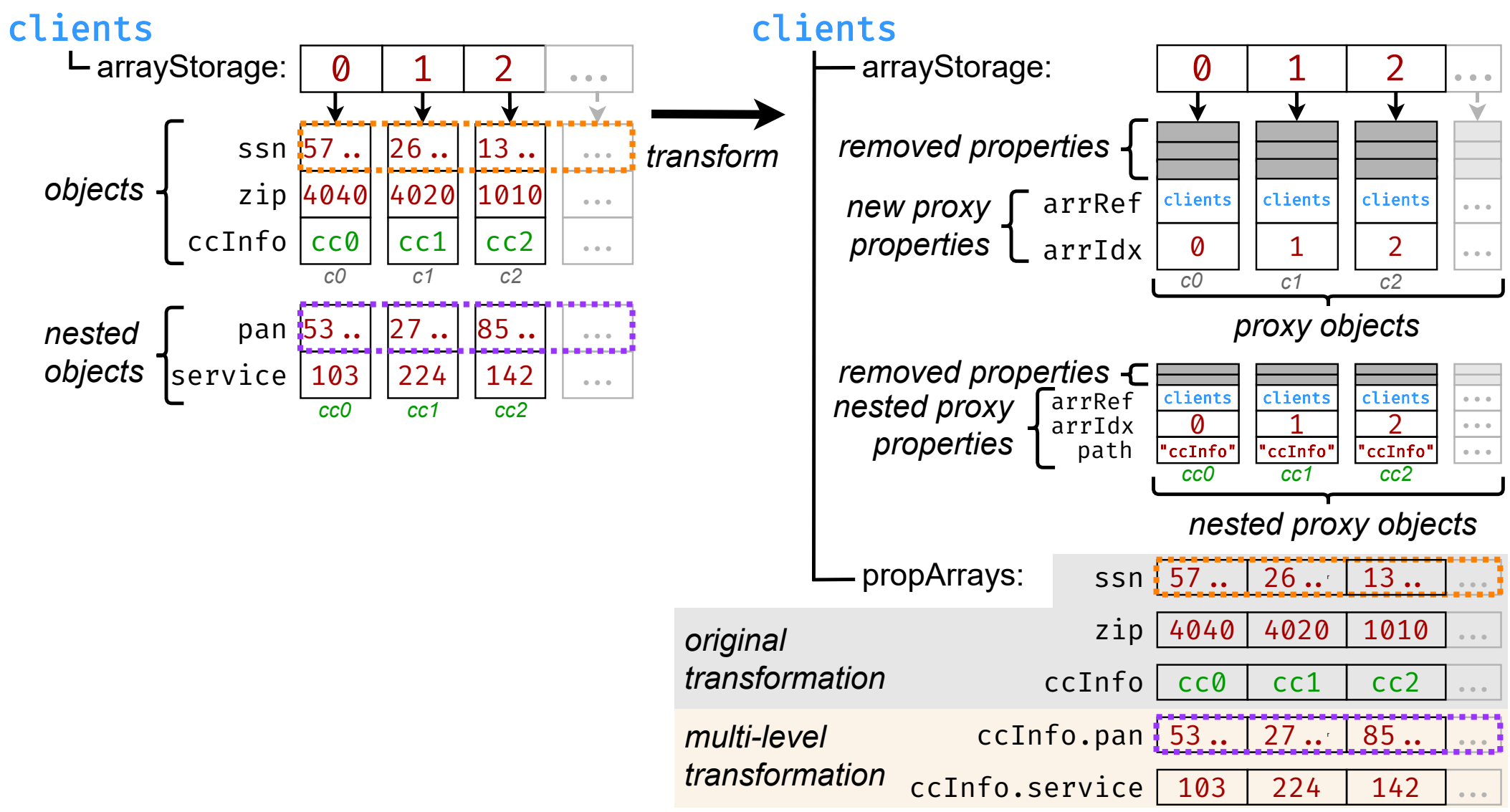

■ **Figure 3** Multi-level transformation of an array of objects to a columnar array.

Transformation is limited to large, often read arrays but adds some initial overhead. However, custom compiler optimizations can amortize the overhead over time by removing redundant checks, compiling methods depending on the array state, and moving loop-invariant code out of hot loops. Thereby, they achieve speedups of up to 9x in microbenchmarks.

## 3 Multi-level Storage Transformation

The results of Makor et al. [55] show significant speedups when executing queries that process non-object-type properties, e.g., boolean or number. However, the results also show that when working with properties of reference types, e.g., dates, strings, or user object types, for which the actual data is stored in properties of these objects, the effect is reduced or performance even degrades. This is because the property array created from a reference type object (e.g., the property array for the ccInfo property seen in Figure 3) only contains the reference to the actual object and not the data that is eventually accessed (e.g., the pan property of the ccInfo object). Usually, many operations on objects require accessing one or multiple of its properties and therefore access memory locations that are scattered throughout the heap when used in loops. However, this is what should be prevented via storage transformation.

To solve this problem, we propose an approach for automatic *multi-level storage transformation* at run time as an improvement to the storage transformation presented by Makor et al. [55]. We adapt the transformation to not only cover the properties of the objects (e.g., ssn, zip, and ccInfo in Figure 3) in the array that is transformed, but also the properties of nested objects (e.g., the pan and service properties of the nested ccInfo objects). Therefore, we do not only place the properties of the array elements in contiguous memory regions but also properties of nested objects.

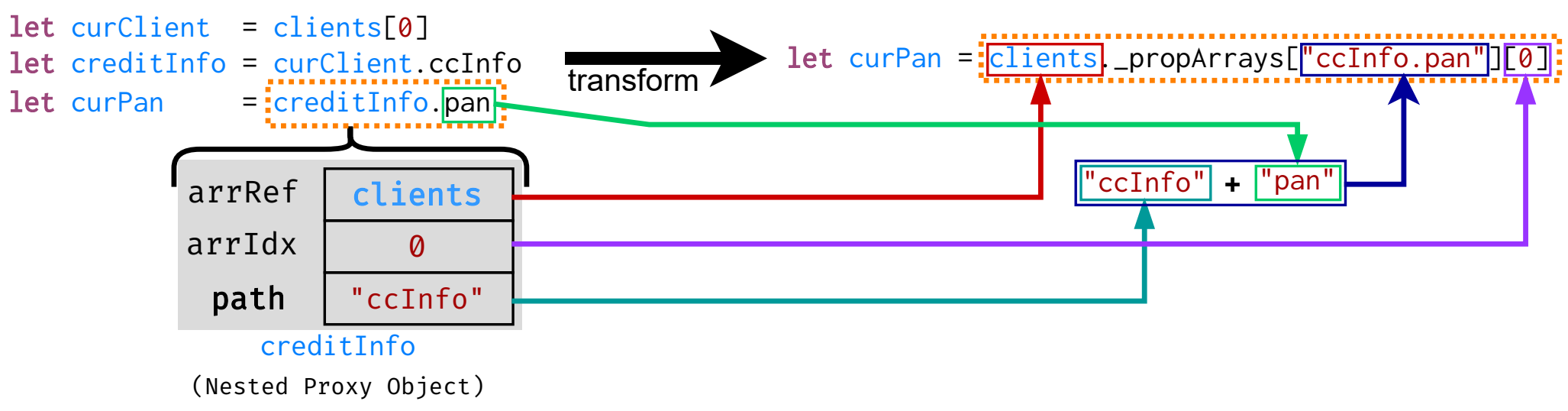

**Figure 4** Accessing a numeric property that is stored in a property array after transformation to columnar storage via a nested proxy object.

An example of applying multi-level storage transformation can be seen in Figure 3. That example is similar to the one presented in Figure 2, except that each client has an additional field ccInfo referencing an object containing the credit card info associated with that client in the form of the primary account number (pan) and the service code (service). When transforming the clients array with the original (i.e., single-level) approach, 3 property arrays for ssn, zip, and ccInfo (containing the references to the credit card info objects) are created as shown in the gray box in the lower right part of Figure 3. However, when performing multi-level storage transformation, in addition to these property arrays, also property arrays for the properties of the nested ccInfo objects are created, i.e., one for the pan property of the ccInfo object and one for the service property of the ccInfo object, which can be seen in the orange box in the right bottom corner of Figure 3. After copying the data of the nested properties into the corresponding property arrays, the properties are removed from the nested objects. Accesses to the properties of the nested objects, e.g., the credit card info objects in Figure 3, need to access these new property arrays. Hence, the nested objects also need to be transformed to proxy objects, similar to how the client objects need to be transformed to proxy objects in [55].

As the objects stored in the array could have multiple nested objects and these objects could have properties with the same name, we cannot store the nested property arrays simply based on the name of the property, but we need to store them based on the path from the array element. In the example depicted in Figure 3 that means that the property array created for the pan property of the nested object referenced by the ccInfo property of the client object is stored using the path "ccInfo.pan". Therefore, this path needs to be known to access the correct property array. When accessing a property only the name of that property is available, as depicted on the left side of Figure 4, where only the name pan is available when accessing the pan property of the creditInfo variable that was loaded from the property ccInfo of the first array element. Unfortunately, as explained earlier, that information is not enough to access the correct property array. Hence, to solve this problem, we not only need to store the arrRef and arrIdx properties for nested proxy objects, but also need to store the property path from the array element to the nested object. An example can be seen in the center right part of Figure 3, where compared to the client proxy objects (which are first-level proxy objects) an additional proxy property, called path containing the string "ccInfo" is stored on the nested proxy objects.

Akin to the original storage transformation, loading property arrays is independent of the nested proxy objects, i.e., independent of the specific client and ccInfo objects in Figure 3, and can typically be moved out of loops when the whole array is processed (cf. compiler optimizations in Section 4). Thus, only the indexed access to the correct property array remains inside the loop, thereby ensuring that subsequent iterations access spatially close memory locations, hence improving the cache utilization.

In Appendix A, we describe how this approach can also handle special cases such as when the nested proxy object is part of different arrays.

**Case Study: Multi-level Storage Transformation for JavaScript Date Objects** We implemented the multi-level storage transformation as described in Section 3, but limited it to Date objects. As JavaScript date objects are internally represented using a timestamp, we adapted the transformation step to perform multi-level storage transformation for nested date objects and thus create a property array for the timestamps of the nested date objects. We focus on date objects, as they are often used in OLAP queries. The heuristics when to apply the transformation remain the same as in [55].

By applying the multi-level storage transformation to Date objects we were able to achieve up to 2x speedups in microbenchmarks that are based on processing Date objects. More detailed results are presented in Section 5.

## 4 Control Flow Duplication for Columnar Arrays

As shown by Makor et al. [55], performance improvements with columnar arrays heavily rely on the optimization of corresponding accesses, particularly in loops.

When compiling code involving columnar arrays, our goal is to generate code, where loops over arrays are optimized towards a specific array state (columnar or non-columnar). We do not want to generate code, where a loop is entered and only then a distinction between columnar and non-columnar arrays is made (e.g., when accessing a property of a proxy object), as this requires additional checks and memory accesses per loop iteration and prevents operations from being moved out of the loop. Unfortunately, this is what a typical compiler would produce without intervention.

Makor et al. [55] rely on *guards* to overcome this problem. Guards are run-time checks embedded into the abstract syntax tree. The GraalVM Compiler subsequently compiles guards to a check that causes deoptimization if the guard fails. Therefore, the code after the failed guard can be omitted from the compilation. Their approach introduces guards that assert a certain array state (i.e., columnar, non-columnar) into methods that access columnar arrays. Similarly, they ensure that any transformation or restoration of an array also causes a deoptimization. Hence, the compiler can optimize such methods (and their loops) with respect to this particular array state.

Guards in GraalVM are generally used in such a way that repeated guard failures cause the compiler to omit the deoptimization and simply compile a more general version of the method that includes both cases (i.e., the one where the guard fails and the one where it succeeds). While this typically makes compilation more complex and may even result in less optimized code (as subsequent instructions that may depended on the guard can no longer be optimized/removed), it increases performance for otherwise frequently recompiled methods, as it prevents excessive deoptimization and allows the methods to stabilize, i.e., recompilations decrease and disappear if no more guards exist or are violated.

To ensure fast access to columnar arrays, the approach in [55] has to prevent this stabilization, as without guards the distinction between columnar arrays and non-columnar arrays would again have to be made on a per array access basis (within loops among others). Despite their evaluation showing that (large) columnar arrays can benefit from this approach, as the run-time costs are outweighed by the gained speedups, particularly smaller arrays or methods that are less frequently called mostly experience slowdowns due to the transformation overhead and frequent recompilations.

We propose an improvement to this approach. When compiling methods, we often distinguish between a regularly used and therefore heavily optimized path—the *fast path*—and a less used path containing more expensive operations—the *slow path*. As an example, we can make this distinction for an addition of two values in JavaScript: `a + b`. By checking the types of $a$ and $b$, we can differentiate between the fast path, where two *numeric* values are summed up, and the slow path, where a more expensive string concatenation of two arbitrary objects is performed.

We can apply the same principle to the compilation of a method with a potentially columnar array: The fast path assumes that the array is columnar and can apply aggressive optimizations. In the slow path, the array is initially not columnar but may be transformed to a columnar variant. Hence, this version has to check the array state upon each access and, therefore, cannot be optimized significantly. Frequently, guards are used to verify whether the fast path can be taken, such that the slow path is only ever compiled if it is indeed necessary for the program. In the context of columnar arrays, this was the approach chosen by [55], where only ever one path (e.g., the columnar path) is compiled to enable optimizations. However, we want to improve upon the aspect that methods that are called with both columnar and non-columnar arrays have to be recompiled per state change. As we want to preserve both paths while still distinctly optimizing the fast path, using guards is not an option.

To solve these problems, we present a new approach with the following goals: *i*) Apply compiler optimizations on columnar array accesses while also preserving accesses to the non-columnar array and *ii*) ensure a stable compilation state with respect to deoptimizations caused by the array state, even for methods that can be invoked with both columnar and non-columnar arrays. Therefore, we decided to use *code duplication* to accomplish our goals.

Code duplication has frequently been used in compilers to enable subsequent optimizations [17, 53, 61, 62]. As it implies a trade-off between the achievable performance gains via those follow-up optimizations and the code size increase, fine-tuned heuristics are often used to influence decisions on whether to duplicate code or not [58].

Figure 5 exemplifies how duplication can enable subsequent optimizations on columnar array accesses while preserving non-optimized code versions to prevent deoptimizations. **(a)** shows accesses to an array of objects (clients). The storage transformation approach shown by Makor et al. [55] extends the language implementation to enable accesses to columnar arrays. In our new approach, we want to preserve all ac-

```
const len = clients.length
const cost = Array(len).fill(0)
for (let i = 0; i < len; i++) {
  const c = clients[i]
  cost[i] = c.zip == 4040 ? 15 : 20
}

// shape of client[i]: {
//   id: <number>,
//   birthdate: <Date>,
//   zip: <number>
// }
```

**(a)** original

```
const len = clients.length
const cost = Array(len).fill(0)
for (let i = 0; i < len; i++) {          <- duplicable loop
  const c = clients[i]                    <- target array
  if (<transitive>(clients,c._isProxy)) {
    $arr = c.arrRef                       } columnar access
    $i = c.arrIdx
    $zip = $arr._propArrays[/*zip*/]
    cost[i] = $zip[$i] == 4040 ? 15 : 20
  } else {
    cost[i] = c.zip == 4040 ? 15 : 20     } generic access
  }
}
```

loop body

**(b)** after storage transformation

```
const len = clients.length
const cost = Array(len).fill(0)
if (<isColumnar>(clients)) {
  // columnar branch (fast path)
  for (let i = 0; i < len; i++) {
    // loop body
  }
} else {
  // generic branch (slow path)
  for (let i = 0; i < len; i++) {
    // loop body
  }
}
```

**(c)** loop duplication

```
const len = clients.length
const cost = Array(len).fill(0)
if (<isColumnar>(clients)) {
  // columnar branch (fast path)
  for (let i = 0; i < len; i++) {
    const c = clients[i]
    if (<transitive>(clients,c._isProxy)) {
      $arr = c.arrRef
      $i = c.arrIdx
      $zip = clients._propArrays[/*zip*/]
      cost[i] = $zip[i] == 4040 ? 15 : 20
    } else { ... }
  }
} else { /*generic branch (slow path)*/ }
```

**(d)** branch optimization

```
// columnar branch (fast path)
$zip = clients._propArrays[/*zip*/]
for (let i = 0; i < len; i++)
  cost[i] = $zip[i] == 4040 ? 15 : 20

        ⇩ loop vectorization

for (let i = 0; i < len; i += L)
  cost[i:i+L] = $zip[i:i+L] ==
    vec<4040xL> ? vec<15xL> : vec<20xL>
// vec<CxL>: vector of constants C with L elements
```

**(e)** loop-invariant code motion

**Figure 5** (Compile-time) Duplication and optimization of columnar accesses (high-level representation)

cess variants to arrays, hence we do not use guards for the array accesses. This results in **(b)**. Parts **(c-e)** depict the steps of this new approach. First, it shows the duplication of the whole loop **(c)**, with a prior check ensuring that one of the resulting loops solely handles accesses to the columnar variant of clients. This distinction enables additional optimizations on the "columnar branch", such as removal of redundant checks **(d)** and finally optimization of the loop itself **(e)**. The following sections highlight individual optimization and preparation steps from Figure 5. Note that aside from Section 4.1 the next sections present new ideas compared to [55].

## 4.1 Columnar Array Markers and Intrinsics

Our approach requires close interaction between the language implementation and the compiler to achieve performance improvements. The runtime needs to relay certain run-time information and profiling data to the compiler to aid detection and optimization of accesses to columnar arrays. At compile time, we need to identify duplicable loops that contain suitable columnar accesses. Therefore, we use a variety of *intrinsics* at the language implementation level. During interpretation, intrinsics merely represent calls to built-in functions. The compiler detects those functions and replaces them with inlined IR or assembly code. In our approach, we also use intrinsics to mark certain accesses and operations, e.g., a transformation of an array to columnar storage. Hence, we also refer to these kinds of intrinsics as *markers*. Table 1 describes some of the custom intrinsics that we use in our approach.

In Figure 5, for example, we use the run-time check <isColumnar> to check the state of the array. At compile time, we detect this intrinsic and can thus infer in which of the corresponding branches the array is known to be columnar. Similarly, we use <transitive> to mark checks which may be optimized for a known array state at compile time. For example, when an object is loaded from an array that is *known* to be columnar, we transitively know that the object *is* a proxy object. Hence, such checks can be removed. To prevent runtime errors, the compiler would usually add numerous additional bounds, type, and null checks to the inflated code for accessing properties in Figure 5 **(b)**. Therefore, we also use special intrinsics for unmanaged accesses to prevent these checks whenever they are redundant, e.g., bounds checks on property arrays that are already safeguarded by the bounds check on the array itself. For brevity, they are omitted from the figure.

**Table 1** Custom intrinsics and markers used to relay information to the compiler

| Intrinsic | Description |
|---|---|
| <isColumnar>(arr) | Checks whether the array is columnar. At compile time, this check is used to identify branches where an array is known to be columnar. |
| <transitive>(arr, cond) | Checks on arrays which are always true for columnar arrays and false for non-columnar arrays. The condition cond can be replaced if the array state is known. |
| <transform>(arr) | Initiates storage transformation on the given array (cf. Section 2.3). The return value indicates whether transformation was successful. |
| <restore>(arr) | Restores the given columnar array to its original form. |

### 4.2 Duplication, Selection, and Optimization of Loops

In our approach, we use duplication on whole loops representing queries that access (potentially) columnar arrays. The overall principle is depicted in the Graal IR shown in Figure 6. The left-hand side of the figure shows the control flow (red arrows) of a simple method containing a loop $l$ with accesses to an array $\alpha$ (blue arrows denote data flow dependencies) that can appear both in a columnar and in a non-columnar state. $A$, $B$, and $C$ denote arbitrary other nodes that appear *before*, *within*, and *after* the loop. The if at the start of the loop denotes the loop condition that decides whether the loop should proceed or end.

As shown on the right-hand side of Figure 6, we duplicate the target loop that contains accesses to a (potentially) columnar array and distinguish between the two duplicates with an <isColumnar> check ensuring that the left loop ($l'$) is only entered if the array is columnar. Subsequently, the loop can be optimized for columnar accesses. As the array might be transformed to columnar storage inside the original loop ($l$), that loop must handle both columnar and non-columnar accesses. Hence, we consider it the *slow path*. After the loops, the control flow merges again.

The GraalVM Compiler already provides means to duplicate parts of the IR [67]. Based on this, our new technique has to *i*) correctly identify a suitable loop that contains columnar array accesses, *ii*) insert a check for the array state before the target loop, *iii*) insert the duplicated loop into the IR, and *iv*) optimize both loops individually based on the determined array state.

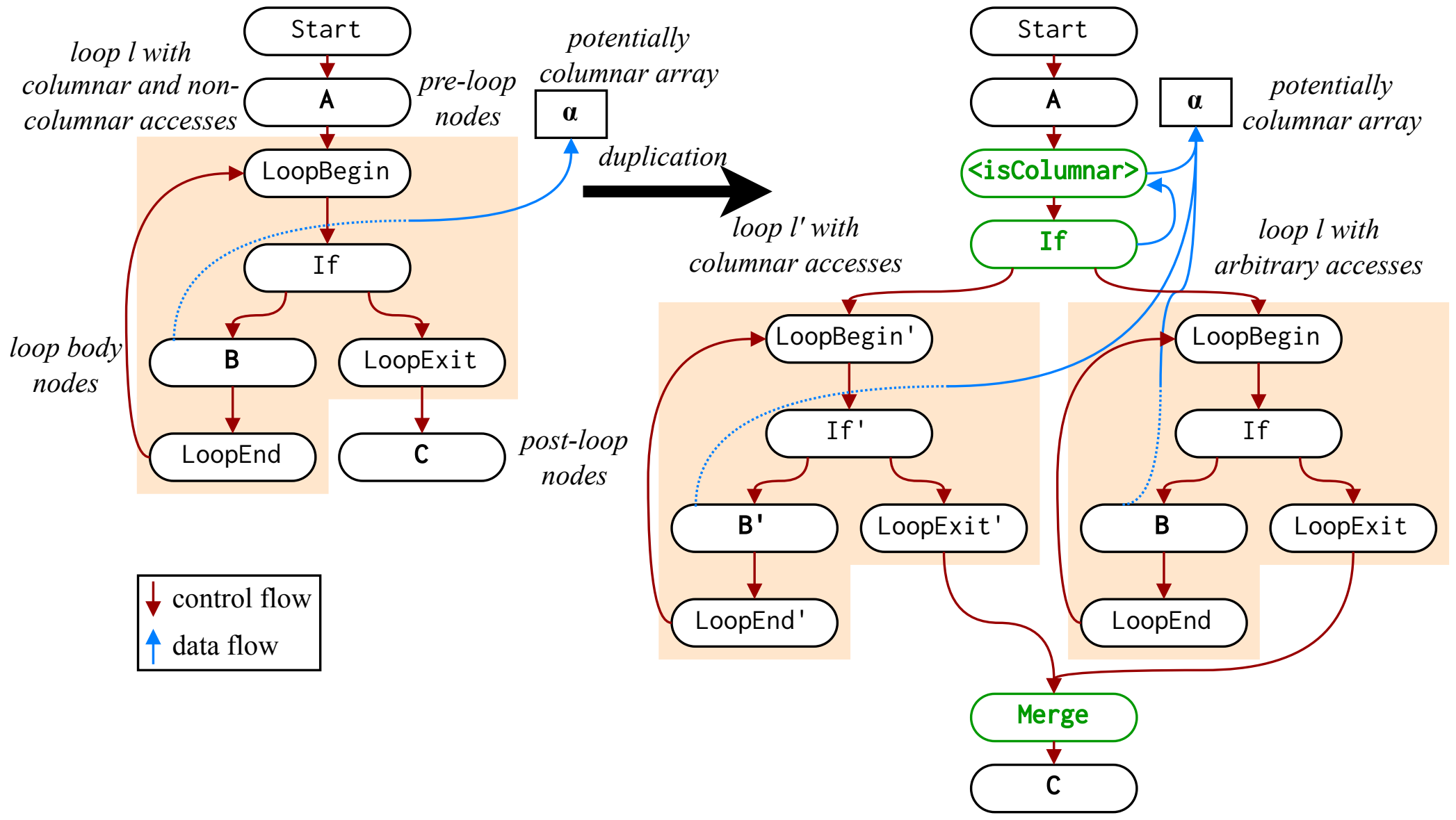

**Figure 6** Control-flow duplication for columnar array accesses

These individual steps are described in detail in the following sections.

#### 4.2.1 Loop Analysis and Selection

As duplication causes a significant code size increase, we added an additional analysis step prior to duplication, to select which loops to duplicate with respect to a target array. This selection process also has to ensure that we can safely optimize the loop. In the language implementation, we ensure that any changes to the array that are not supported by the columnar format cause a restoration. We also use the intrinsics from Section 4.1 to mark array accesses, transformations, and restorations. This allows us to identify them at compile time and accumulate metrics for each loop and array. In summary, our duplication approach has the following requirements for a loop:

1. The loop must contain *read* accesses to a potentially columnar array.
2. The loop must have been executed with the target array in a columnar state.
3. The loop must not contain array restorations or operations that may cause a restoration (e.g., element insertions, deletions). However, the loop may contain array transformations from non-columnar to columnar format.
4. The loop must not contain method calls or other nodes which can allow the array to escape (and potentially change its state). While this may appear limiting, the GraalVM partial evaluator inlines most AST operations and nested calls, such that they should only appear infrequently within loop bodies [42, 50, 87, 89].

#### 4.2.2 Loop Duplication

Before duplication, we insert a custom <isColumnar> check as shown in the example in Figure 5 **(c)** and in the IR in Figure 6. This and the knowledge that the columnar array is never restored to its original state within the loop (cf. Section 4.2.1) ensures that on the branch optimized for columnar access we can safely assume that the array is strictly columnar. Therefore, we refer to this branch and the corresponding loop as the *columnar branch*. We call the other branch and its loop the *generic branch*, as it contains both columnar and non-columnar accesses.

Duplication of the loop nodes is already supported by the GraalVM Compiler. This process also *merges* the control flow after the loops and introduces $\phi$ nodes for values that depend on the loop that was used. Hence, the IR nodes after the merge are not affected by the duplication. Finally, we attach the duplicated loop to the corresponding branch of the <isColumnar> check.

#### 4.2.3 Optimization of Duplicated Loops

The main goal of our approach is the optimization of the columnar branch. As stated before, we know that the array is columnar throughout this version of the loop. Hence, after duplication, we can remove redundant instructions, such as array state checks (marked by the <isColumnar> intrinsic) as well as other checks that are transitively true for columnar arrays (<transitive>), e.g., checking whether an object loaded from the array is indeed a proxy (cf. Figure 5 **(d)**). As the generic branch is not limited to a single array state, it cannot be optimized significantly at this point. However, in Section 4.3 we describe an improvement that also enables optimization of this branch to a certain degree.

**Loop-Invariant Code Motion** *Loop-invariant code motion* [5] is a compiler optimization already utilized by the GraalVM Compiler that attempts to move instructions from within the loop body to before the loop if they are not affected by the execution of the loop. Hence, it prevents redundant memory accesses and computations that can be performed once prior to the loop instead of in every loop iteration. With the aforementioned optimizations, we can make most of the overhead for accessing columnar arrays loop-invariant. Thus, the majority of operations can be moved out of the columnar loop, such that often only the property array accesses (i.e., the accesses to the actual data as shown in Figure 5 **(e)**) remain.

**Enabling Loop Vectorization** Since replacing field accesses with indexed accesses is an inherent feature of columnar arrays, accesses to such data structures lend themselves to loop vectorization [15]. Loop vectorization describes the process of optimizing loops by using hardware-supported SIMD instructions [43, 44]. By combining individual loop iterations, certain patterns in loop bodies containing arithmetic operations or conditions can be rewritten, such that multiple operations can be performed in parallel (e.g., eight 32-bit integer values (placed contiguously in memory) can be modified/loaded/written at once by using a single 256-bit vector instruction).

As the GraalVM Compiler already supports loop vectorization [24], the challenge for us is to ensure that loops involving columnar arrays are in a proper shape, such that they are subsequently picked up by the vectorizer. As mentioned before, our custom compiler phases optimize the loop body such that often only the property array accesses remain. Hence, for queries such as those in Figure 5 **(e)**, some operations on property arrays can be replaced with SIMD instructions, resulting in additional performance benefits (cf. Section 5.1).

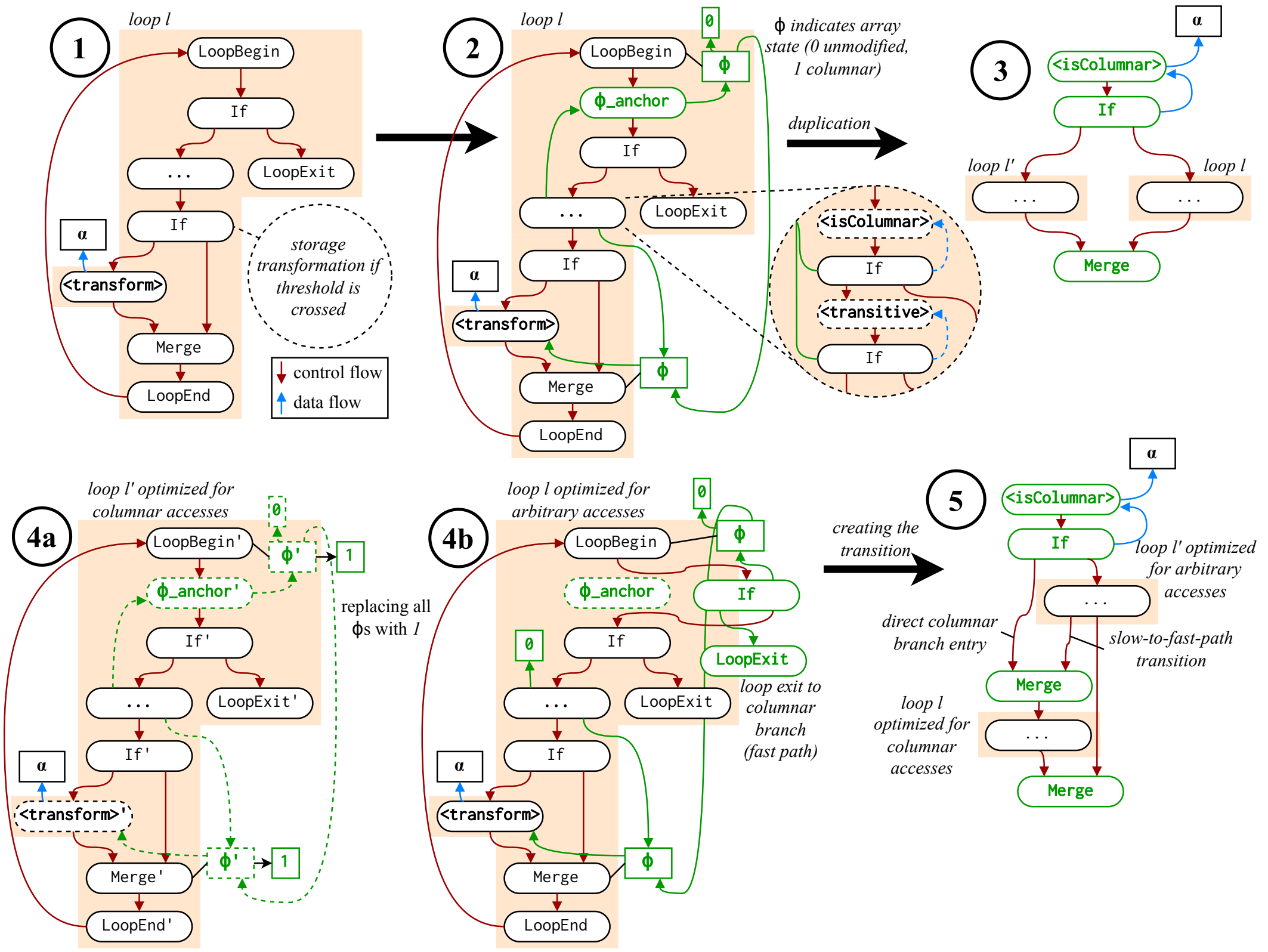

**Figure 7** Slow-to-fast-path transition and generic branch optimization

### 4.3 Refinement: Slow-to-Fast-Path Transition

One problem with our duplication approach so far is that it does not provide benefits if the optimized method is mostly called with arrays that are only transformed in loops of the method itself. Consider duplication in a method consisting of a single loop similar to the one on the right-hand side of Figure 6: If the method in question is called with an array *not yet* in columnar shape, the generic branch is taken. In the loop, the array may be transformed into a columnar layout if the access threshold is crossed, however, execution still resumes in the generic loop. Since optimization is limited to the columnar branch, the array never benefits from the duplication as the entirety of the execution is spent in the generic branch even after transformation of the array.

Based on this observation, we implemented a refinement of our duplication approach that allows *a*) transitioning to the columnar (fast) branch if the array is transformed while executing in the generic branch and $b$) some limited but effective optimization of the generic branch itself.

Figure 7 illustrates the main steps in this new duplication approach:

1. The analysis step from Section 4.2.1 selects and prioritizes suitable loops based on gathered array metrics and the shape of the loop.
2. We add a new loop $\phi$ node. Its binary value represents the state of the target array at every point in the loop (0 for non-columnar, 1 for columnar). We attach this new $\phi$ node to a custom node in the control flow we call the $\phi$ *anchor*. This node serves no other purpose than to mark the beginning of the loop and capture the $\phi$ value at the start of a loop iteration. It is removed in a later step of this process. We then iterate the nodes of the loop in *dominator order* (cf. Section 2.1), always tracking the last known state of the array.
   a) We replace custom checking intrinsic (e.g., <check>, <transitive>; cf. Section 4.1) on the *same* array with the last known state.
   b) When encountering a transformation call (<transform>) for the array, we use the return value of this function as the new array state (i.e., if the transformation succeeded, the return value is 1, otherwise 0).
   c) At merges, we create additional $\phi$ nodes that combine the incoming states, in case there was a transformation in one or multiple of the paths.
   d) At every loop end, the last state is propagated to the initial loop $\phi$.
3. We duplicate the loop and insert the preemptive check and branching condition.
4. The resulting branches are optimized separately:
   a) In the columnar branch loop $l'$, all $\phi$ nodes representing the array state are replaced by a constant 1. Afterwards, a built-in optimization phase optimizes the now constant conditions, removes redundant branches, etc.
   b) In the generic branch loop $l$, we replace the $\phi$ anchor with a new if condition that checks the value of the $\phi$ node at this point and exits the loop if the array has become columnar. Otherwise (the $\phi$ value is 0), the generic loop is resumed. In that case, any usages of the $\phi$ anchor are replaced with a constant 0, thus signaling that the array cannot be columnar if the loop is continued. Subsequently, constant conditions can be optimized. The remaining $\phi$ nodes of merges in the loop denote cases where the array state is still uncertain.
5. After individual optimizations, the newly introduced loop exit from the generic loop is merged with the point before the loop header in the columnar branch, thus enabling a transition from the generic branch to the columnar one. Additionally, the existing loop $\phi$ nodes of the columnar loop have to be patched to reuse the accumulated $\phi$ values from the generic branch if it was executed before (e.g., if a counter is used, the columnar loop should reuse the last counter value from the generic branch and not reset it to the initial value).

The algorithm works similarly for loops where multiple arrays appear in columnar format. In this case, we track the state of all these arrays and only allow the slow-to-fast-path transition if *all* arrays are columnar.

## 5 Evaluation

In comparison to the work done by Makor et al. [55], we want to improve the performance especially on smaller workloads with fewer iterations by limiting the recompilation overhead. Our multi-level storage transformation approach should yield performance benefits for arrays of objects with Date properties. We also want to introduce no significant overhead for workloads not suited for our approach.

As our approach is integrated into the GraalVM JavaScript runtime, we also have to ensure compliance with the standard. Our approach passes the V8 and ECMAScript test suites [34], which in combination consist of over 51000 test cases. During those runs, we lowered the threshold of the transformation significantly to force storage transformations of smaller arrays. Overall, around 450000 transformations were attempted on arrays, of which around 1000 succeeded. However, around 750 of those arrays had to be restored later as unsupported operations occurred.

Unfortunately, data-heavy benchmarks over large arrays are lacking in JavaScript and traditional JavaScript benchmark suites—e.g., Octane [16], JetStream [36]—typically do not contain workloads that we target with our approach. Hence, we reused and extended the microbenchmarks[1] and TPC-H [23] ports from [55].

In Appendix D, we evaluated the slow-to-fast-path transition introduced in Section 4.3. In Appendix E, we evaluated the overhead of the transformation itself. In Appendix F, we also show that our approach causes no significant overhead on workloads that do not contain applicable arrays. The approach presented in this paper is based on a newer version of GraalVM, hence we had to port the approach by Makor et al. [55] to make a valid comparison. Therefore, the achieved benchmark scores may slightly differ from their evaluation. In all benchmarks, we use a minimum array size of 5000 and a minimum read count of 2500 for an array to be eligible for transformation. These thresholds are lower compared to the original version in [55] as the duplication approach enables faster improvements for smaller arrays.

Benchmarks—unless otherwise noted—were executed on a dual-socket Intel(R) Xeon(R) CPU E5-2690 @ 8 × 2.90 GHz (32 logical cores, 192 G available RAM). Our approach is based on GraalVM EE version 22.3. To ensure appropriate warm-up, we repeated each benchmark several times before measurement.

### 5.1 Microbenchmarks

Figure 8 shows the performance of our approach on a set of microbenchmarks compared with [55] and GraalVM JavaScript without storage transformation. Each benchmark executes a particular query on a large array of objects. The depicted results vary in the number of times the query is applied to the same array (100, 1000) and in the sizes of the generated arrays (10k, 1M). Appendix B contains results on additional configurations and a more detailed description of the individual microbenchmarks. The results emphasize the key benefit of duplication: Compared to [55], our approach now already achieves speedups on the smallest presented configuration with only few iterations on certain queries (around 2x on *aggregate* and *writeProperty* with 100 iterations). The date/string variants of filter/aggregate contain workloads with nested objects. The results show how our multi-level storage transformation on dates enables speedups for the queries involving dates. With larger arrays and more iterations, aggregate and writeProperty achieve speedups of around 10x and 7x, compared to the 8x and 6x from the approach presented in [55]. The columnar variant of *mapClientsToCosts* is SIMD vectorized and shows a speedup of up to 2x (cf. Section 4.2.3).

[1] Code: https://github.com/lmakor-jku/data-intensive-js-benchmarks (visited on 2022-12-30).

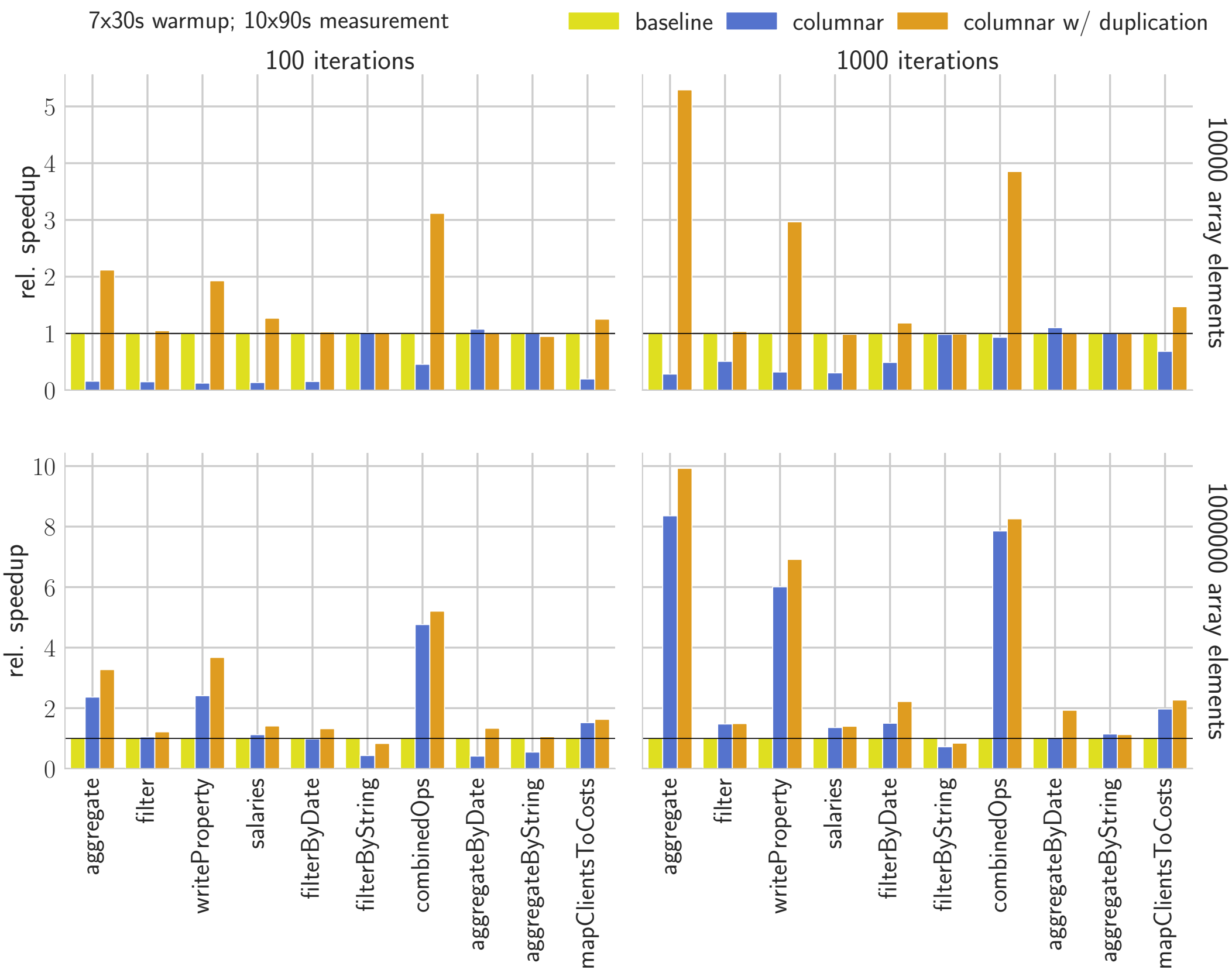

**Figure 8** Microbenchmark throughput of our approach compared to [55] and GraalVM JavaScript without storage transformation *(higher is better)*

### 5.2 TPC-H

The TPC-H Decision Support Benchmarks [23] are a benchmark suite frequently used to measure database performance. Containing a variety of queries over a large dataset, these benchmarks seem suitable for also evaluating our approach. Unfortunately, they are only available in SQL format (a problem also noted by [56]), hence we chose to use the same custom JavaScript ports as in [55]. These ports contain 5 of the 22 queries that are part of TPC-H: *q1*, *q6*, *q12*, *q14*, and *q19*. The selection of these particular queries is detailed in Appendix C and was based on the individual query properties to show performance of our approach using different query operations and features.

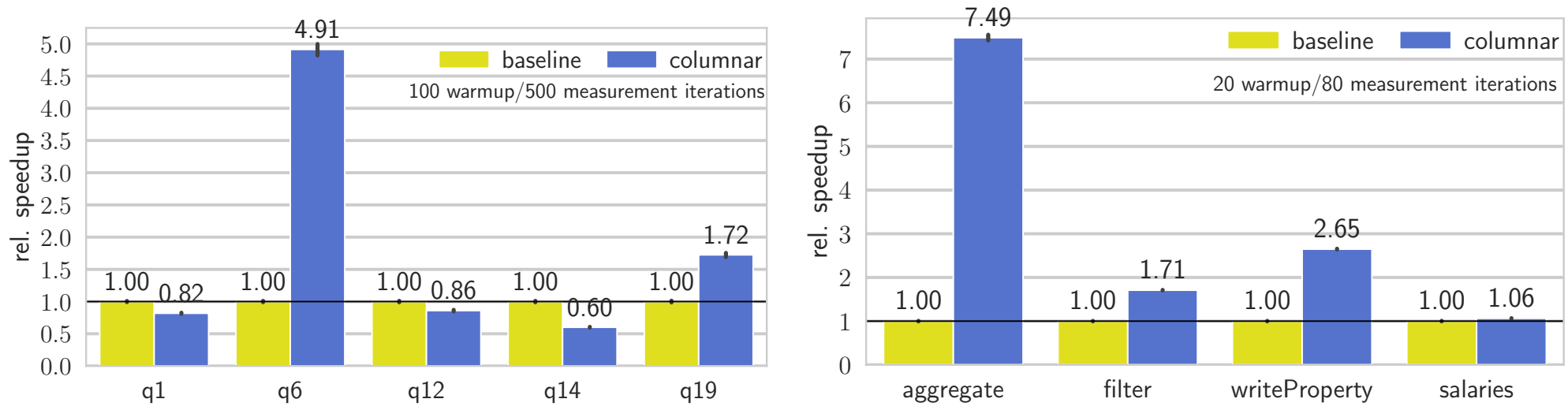

**(a)** TPC-H **(b)** Lodash

■ **Figure 9** Query performance of different benchmarks with our approach relative to baseline without storage transformation *(higher is better)*

■ **Table 2** perf cache-miss evaluation

| Metric | Baseline Count | Columnar Count | Reduction |
|---|---|---|---|
| cache-misses | 2 575 530 996 544.6 | 50 406 360 352.9 | 51.1x |
| L1-dcache-load-misses | 3 142 811 326 847.5 | 137 072 211 496.1 | 22.9x |
| LLC-load-misses | 520 871 505 214.4 | 530 809 018.5 | 981.3x |

Figure 9a shows the results of this evaluation. Queries *q1*, *q12*, and *q14* perform poorly on our approach compared to their execution on standard GraalVM JavaScript. This is due to their increased usage of grouping and string functions which are not affected by our optimizations. Additionally, they frequently contain operations that are not perfectly inlined by GraalVM's partial evaluator, thus resulting in method calls that prevent our duplication from considering this query. Queries *q6* and *q19*, however, are improved, with *q6* experiencing a major performance increase (nearly 5x).

## 5.3 Lodash

For the results in Figure 9b, we executed our approach on a number of Lodash [26] queries based on microbenchmarks (cf. Section 5.1) and compared the performance with the execution on standard GraalVM JavaScript. As depicted, our approach causes significant speedup for most queries, with *salaries* receiving only slight improvements.

## 5.4 Cache Behavior

We evaluated the cache behavior of our approach compared to GraalVM JavaScript without storage transformation using *perf stat*. We executed the aggregate microbenchmark with an array size of 5M and 4000 iterations to ensure that the compiler threads do not impact the counters. One of the main goals of the columnar layout is the better data locality due to the property arrays. Hence, we expect applications using arrays with columnar layouts to exhibit less cache misses than ones that use common arrays. Table 2 contains the results of this evaluation for different *perf* events. The columnar variant shows significantly less cache misses. Due to hardware constraints, this evaluation was conducted on a single-socket Intel(R) Core(TM) i7-1065G7 CPU @ 1.30 GHz with 8 logical cores, 32 G available RAM, and 4 × 192 KiB L1d, 4 × 128 KiB L1i, 4 × 2 MiB L2, and 1 × 8 MiB L3 cache.

## 6 Limitations

Applying loop duplication induces costs in terms of both compile-time and code size. Hence, query duplication should only be applied to queries that can be optimized through duplication. Similarly, applying multi-level storage transformation requires careful consideration, as it incurs even more transformation overhead.

**Query Analysis** Our loop analysis selects loops with columnar array accesses that do not restore the array to its original format. A more sophisticated analysis, such as detecting the kind of the query (e.g. update, mapping, or filtering query), could allow more specific optimizations. This decision has to consider that unoptimized access to columnar storage is significantly slower than common property access.

**General Multi-level Storage Transformation** So far, we implemented multi-level storage transformation only for `Date` objects. Hence, one additional level of transformation has to be applied at most. Multi-level storage transformation for arbitrary types would mean that arbitrary levels of nested objects could be present and would need to be handled. Besides, general multi-level transformation would have to consider a potentially circular object hierarchy. One solution for that could be a limit on the number of levels that should be transformed.

## 7 Related Work

We group research that is related to our work into three distinct categories: a) Research on columnar and memory-layout-focused data structures, b) query optimization techniques, c) and related compiler optimizations.

### 7.1 Columnar Data Structures

While there are several approaches that integrate columnar data structures into common language runtimes, the majority of those operate on existing columnar data or provide APIs for such data structures. Due to the static data layout, no *run-time* transformation has to occur, hence developers can leverage these approaches specifically for high-performance applications. It also has the benefit that compilers can often automatically optimize accesses to such collections as the data layout does not change drastically at run time. The downside of these approaches compared to our work is that they require developer intervention to use columnar storages. To the best of our knowledge, there is no contemporary work proposing automatic *run-time* storage transformation as we do.

Data analysis and visualization libraries such as *Apache Arrow* [7] and *RAPIDS* [75] also support columnar data structures for JavaScript and other dynamic languages. *DataFrames* allow tabular layouts in Python [80] and Julia [46]. Julia also supports a *StructArray* type that provides an abstraction over a struct of arrays [45].

Mattis et al. [56] implemented a columnar API for Python. Similar to our approach, they represent elements of columnar arrays as *proxy objects*. The JIT compiler of the PyPy runtime can leverage the columnar data layout to improve performance without modifications to the compiler. They achieve that by limiting the lifespan of their proxies by emitting new ones upon element access. This allows the JIT compiler to prevent proxy allocations altogether in certain cases.

Pivarski et al. [71] perform code transformations on the AST level to enable fast access to an *already columnar* data set. This enables access to such data via traditional object-oriented APIs and avoids materialization of objects from the originally columnar data. We share a common goal in shifting the responsibility of optimizing data accesses to the runtime and compiler. However, their data is already in columnar form and they do not target an automatic transformation of non-columnar data.

Noth [64] introduced *exploded* Java objects, where field values are stored in parallel arrays to reduce the memory footprint and improve performance via better cache utilization. Exploded objects resemble our proxy objects, with the important difference that proxies are always part of a (columnar) array, while exploded objects are autonomous. Thus, the field values of exploded objects are stored in global, parallel arrays instead of per-array property arrays. In contrast to our approach, they perform a Java source-to-source translation, with users annotating classes that produce exploded objects. This allows them to support most Java semantics on exploded objects without modifications to the JVM. Our approach requires modifications to both the language implementation and the compiler but enables automated transformation of arrays.

De Wael et al. [28] implemented an API for *just-in-time data structures*. Such data structures are defined by the developer via a custom DSL and contain transformation rules that enable changing the underlying data layout at run time. While this means that developers have to be aware of optimal data layouts for certain scenarios, they also propose an approach that utilizes machine learning to automatically select the best data layout based on execution time.

### 7.2 Query Optimization

Zhang et al. [90] generate efficient query execution plans for loops in user code that are marked by so-called *pragmas*. They use the Truffle framework to generate an AST from the loop and subsequently apply different query plans. These are evaluated using both online and offline analysis to select the ones that perform best. As we apply query optimizations during compilation, we do not plan adding additional evaluation steps in order to not impact compilation time further. In our approach, users do not have to mark or preselect queries that should be optimized; instead, such queries are automatically optimized if columnar arrays are used.

Shaikhha et al. [76] pursue a different direction, where they take an incoming query aimed at a columnar store and subsequently perform compiler-like optimizations in a high-level language. This includes different strategies for materializing the corresponding rows from the columnar store. Many of their proposed query optimizations are already implicitly performed by the GraalVM Compiler. Regarding materialization, we seem to have similar goals, as they try to prevent unnecessary materialization of rows while we try to remove redundant array element accesses. Chen et al. [18] propose a similar approach, where they perform a variety of compiler optimizations on an IR generated from database queries. Grulich et al. [40] also perform optimizations on an IR but support polyglot execution of queries mixed with user-defined functions via the Truffle framework. As the input to all of these approaches are dedicated queries, they can apply more specific optimizations depending on the query types. Conversely, they do not aim at changing the underlying data store—a pivotal step in our approach.

### 7.3 Compiler Optimizations

Using duplication in a compiler is typically a trade-off between code size and the expected gains of possible follow-up optimizations. While the concept itself has already been successfully applied over 20 years ago [61, 62], there is also more recent work that somewhat relates to our approach.

Zhao et al. [91] use duplication to parallelize loops in a JVM. Their detection of suitable loops is similar to our loop analysis, although with different aims and filter criteria: They inspect the shape of a loop, i.e., the exit criterion and parameter shapes, while we use intrinsics introduced into the language implementation to find targets. Similarly, they add a prior check to their duplicated loops that enables optimizations such as bounds check removal in one loop. As they consider static properties of arrays (i.e., their length), they do not have to worry about intermediary changes. We check the array state before loops to make a distinction but this state may change within the loop. Hence, we introduced a *slow-to-fast-path transition* (cf. Section 4.3).

Leopoldseder et al. [53] use *simulation* in the GraalVM Compiler to predict benefits and costs of duplicating code after control flow merges into the individual branches. So far, our duplication decisions are purely based on the aforementioned loop analysis steps. However, we think that using simulation may further benefit our approach to also assess the impact of the duplication and subsequent optimization.

Loop optimizations are prevalent in modern compilers such as PyPy [8, 9, 72], V8 [82], the HotSpot C2 Server Compiler [68], and the GraalVM Compiler [24, 52]. Techniques such as loop peeling [10] and loop unrolling [27] utilize duplication to replicate the loop body. They simplify the loop by unwinding individual loop iterations or enable follow-up optimizations such as parallelization.

In our approach, the duplication guarantees two distinct loops, where each variant handles arrays in a particular state. In contrast to traditional loop optimizations, this process takes into account whether and how a columnar array is modified or accessed in the loop by specifically querying intrinsics from the language implementation. As our columnar arrays are only generated at run-time and cannot be determined statically (i.e., the compiler cannot automatically infer that an array is columnar), this is a necessary requirement to narrow down the set of candidates for a duplication.

Optimizations in our approach similarly are driven by intrinsics, as they highlight optimization potential otherwise inaccessible to the compiler, such as removing checks in the context of a known columnar array and decouple property accesses from array element accesses. Loop-invariant code motion [5] subsequently moves many of the remaining operations out of the loop. Ideally, the columnar loop variant is subsequently comparable to the compilation of a loop over a static columnar array, while the other variant contains the generic accesses to the array (columnar or non-columnar).

The meta-tracing Python JIT compiler PyPy [81] produces so-called *bridges* [8] upon repeated guard failures. This entails a compilation of the trace from the failing guard that may eventually jump back to the compiled trace containing the guard. Using columnar arrays, this could mean that if the state of an array changes frequently (and thus causes guard failures), we create a bridge with the new state. Depending on the applied optimizations and the length of the bridge, the resulting compiled traces may resemble the branches in our duplication approach. However, as the GraalVM Compiler is method-based, (meta-)tracing JIT compiler techniques are not applicable.

## 8 Conclusion

In this paper, we presented an approach for detecting loops representing query-like operations on potentially columnar arrays and optimizing such loops via duplication in a state-of-the-art JIT compiler. Thereby, the duplicated loops are optimized individually, i.e., the generic loop (slow path) that handles access to both columnar and non-columnar storage formats and the duplicated loop that only contains accesses to the columnar storage (fast path). As arrays in our approach may be transformed to columnar arrays during query execution, we furthermore implemented a slow-to-fast-path transition that allows such arrays to also profit from optimizations to columnar accesses. To speed up queries working with reference type properties, we introduced a multi-level storage transformation that not only stores the properties of the objects in columnar form, but stores the whole object hierarchy in property arrays. We implemented this technique as a case study for Date properties.

An evaluation on microbenchmarks and ports of widely used database queries shows that our approach can indeed improve performance of queries on columnar arrays, while still suffering from issues when “grouping” operations are used. Our approach also significantly improves cache behavior for suitable queries. Additionally, our results reveal that multi-level storage transformation can also improve queries using operations such as Date comparisons. Overall, our work could inspire researchers to pursue different ways of using columnar data structures in programming languages.

**Acknowledgements** Authors Sebastian Kloibhofer and Lukas Makor contributed equally to the paper. This research project was partially funded by Oracle Labs. We thank all members of the Virtual Machine Research Group at Oracle Labs. We also thank all researchers at the Johannes Kepler University Linz’s Institute for System Software for their support of and valuable feedback on our work.

## A Multi-Level Storage Transformation Special Cases

To preserve program semantics, multi-level storage transformation as introduced in Section 3 has to be able to handle several different special cases. These revolve around the fact that the transformed nested object properties may reference objects that could also be referenced from elsewhere.

**One object with a nested object in one array** This is the default case where one nested object (e.g., a date) is associated with a single object that is part of an array. After transformation, the nested proxy object has to remember the property name that references it (e.g., "lastChanged") to get the identifier of the property array, the index of the object it is part of, and the transformed array itself. This triple then enables access to property values of the nested object by dereferencing the array and accessing the property array with the given name at the given index position.

**One object with the same nested object in multiple properties** The same nested object could also be referenced by one object via multiple different properties (e.g., `lastChanged` and `deletedAt` properties that reference the same date object). In this case, the transformation would add the values of the one object to two distinct property arrays, one for each property. After transformation, we have to ensure that any writes to one of these property arrays also updates the other (e.g., the update of the `lastChanged.timestamp` property array also has to be applied to the `deletedAt.timestamp` property array).

**One object with a nested object in multiple (columnar/non-columnar) arrays** If an object with a nested object property occurs in multiple arrays, writes to a property array associated with the nested proxy object have to be replicated across all other columnar arrays. For non-columnar arrays, we already ensure that accesses to the nested object properties are correctly rewired to access one of the property arrays by using the aforementioned triple of array, index, and property name.

**One object with a nested object multiple times in the same array** Similarly, one object with a nested object property could occur multiple times in *the same* array at different index positions. There, writes to a property array associated with the nested proxy object have to be reflected onto every index position—e.g., a write to the `lastChanged.timestamp` property at index `i` must also change the value at index `j` if the array contains the same object at both positions `i` and `j`.

**Multiple objects with the same nested object multiple times in the same array** Finally, multiple different objects in the same array could still reference the same nested object. This is a more generic variant of the one mentioned before, where, similarly, writes to one position of a property array of the nested object may also have to be performed to other positions.

Makor et al. [55] already describe how the transformation of arrays into columnar format can cause problems if objects are part of multiple arrays among others. They solved their problems by associating each transformed proxy object with a *set* of array-index pairs, where each pair identifies one location from which the corresponding proxy (or its properties) might be accessed. We adapted this solution to handle the above mentioned cases. However, nested proxy objects require three pieces of information for access to the property array: The associated columnar array that stores the property arrays, the property name referencing the nested object that serves as the property array identifier, and the index of the object in the columnar array that also identifies the position of the nested object property values in the property arrays. Therefore, each nested proxy object has to manage a set of *triples*. Upon write access to a property of a nested proxy object, this set is iterated to perform the write on *every* associated property array. Hence, writes are expensive as soon as a nested object is referenced multiple times.

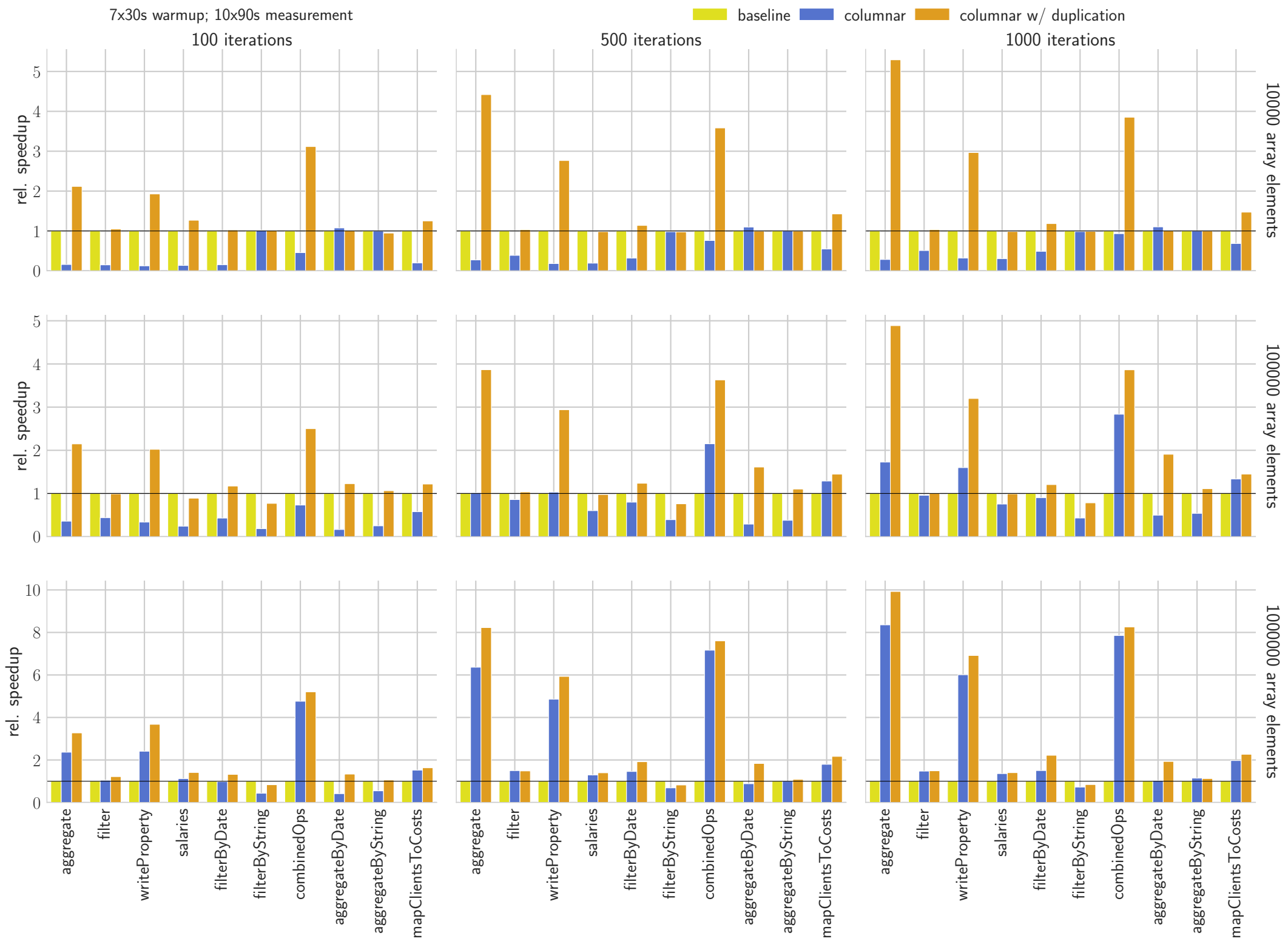

**Figure 10** Microbenchmark throughput of our approach compared to [55] and GraalVM JavaScript without storage transformation *(higher is better)*

## B Microbenchmark Details

Figure 10 contains the results of our microbenchmarks on 9 different configurations. As described in Section 5.1, duplication allows speedups on smaller arrays with fewer iterations but also boosts overall performance on most benchmarks.

**Table 3** Description of our microbenchmarks

| Query | Explanation |
|---|---|
| *aggregate* | Summing the values of a single property. |
| *filter* | Filtering an array based on a single property. |
| *writeProperty* | Computing and updating the values of a single property based on another property. |
| *salaries* | Computing and updating the bonus and salary of an employee based on the achieved salary, while also accumulating all the new salaries. |
| *filterByDate* | Filtering based on a Date comparison. |
| *filterByString* | Filtering based on a string comparison. |
| *combinedOps* | Consecutively performing the *aggregate*, *filter* and *writeProperty* operations. |
| *aggregateByDate* | Summing the values of a single property after filtering based on a Date comparison. |
| *aggregateViaStringEquals* | Summing the values of a single property after filtering based on a string comparison. |
| *mapClientsToCosts* | Calculating the shipping costs based on client ZIP code. Can be vectorized in columnar format. |

The set of microbenchmarks we use for evaluation by now encompasses 10 different queries that should showcase different operations over large arrays. All microbenchmarks share the structure presented in Listing 1: An array is generated with a given size and filled with uniform objects. Note that different benchmarks may use objects with different properties (e.g., filterByString uses objects with string properties in contrast to aggregate that only uses objects with primitive properties). Then, a given QUERY (e.g., aggregate, writeProperty) is repeatedly executed on the array. We ensure that neither the intermediary results nor the array itself can be discarded by the compiler. Table 3 describes the individual queries in more detail.

**Listing 1** Basic template for our microbenchmarks

```
function benchmark(arraySize, queryApplications) {
  // create an array of uniform but randomized objects
  var array = createArray(arraySize);
  // repeat the query on the given array a number of times
  for (let i = 0; i < queryApplications; i++) {
    let result = QUERY(array);
  }
}
```

**Table 4** A subset of the properties of individual TPC-H queries

| Query | GROUP BY | ORDER BY | AGG | SUBSELECT | STRING OP | DATE OP |
|---|---|---|---|---|---|---|
| *q1* | x | x | x | | | x |
| *q2* | | x | x* | x | x | |
| *q3* | x | x | x | | x | x |
| *q4* | x | x | x | x | | x |
| *q5* | x | x | x | | x | x |
| *q6* | | | x | | | x |
| *q7* | x | x | x | x | x* | x* |
| *q8* | x | x | x | x | x | x* |
| *q9* | x | x | x | x | x* | x |
| *q10* | x | x | x | | x | x |
| *q11* | x** | x | x | x | x | |
| *q12* | x | x | x | | x | x |
| *q13* | x | x | x | x | x* | |
| *q14* | | | x | | x | x |
| *q15**** | | x | x* | x | | x* |
| *q16* | x | x | x | x | x | |
| *q17* | | | x | x | x | |
| *q18* | x | x | x | x | x | x |
| *q19* | | | x | | x | |
| *q20* | | x | x* | x | x | x* |
| *q21* | x | x | x | x | x | x |
| *q22* | x | x | x | x | | x |

* has property in subselect or VIEW
** includes HAVING clause
*** creates VIEW
DATE/STRING OP include extraction of parts (extract(year...)) and sorting

## C TPC-H Query Selection

As mentioned in Section 5.2, we evaluated our approach on JavaScript ports of a subset of the TPC-H queries. This selection was done purposefully to evaluate our approach on a set of queries with different properties. The ports themselves only use traditional JavaScript constructs without using external libraries. Our approach does not perform well for more complex queries involving grouping or string operations, as those often result in complex loops and nested method calls that prevent duplication.

Table 4 contains a high-level analysis of *all* TPC-H queries (note that there may be slight differences between variants of individual queries), denoting whether certain queries use certain query features (a more detailed analysis of the TPC-H query set was performed by Boncz et al. [12] who similarly highlight the extensive use of grouping and aggregation). The table furthermore highlights those queries that were selected for the evaluation (*red* where our approach performs poorly, *green* where the approach improves performance). *q6* and *q19* are among the simplest queries in terms of complexity and neither contain grouping or ordering functions, nor use subselects or (complex) string operations—in contrast to queries *q1* and *q12* that encompass a variety of different query features. While *q14* also seems relatively simple, it contains a string operation using the LIKE operator, hence requiring a more sophisticated string comparison than *q19* that only checks for a full match. This difference is significant as the comparison in *q14* results in a vastly more complex IR at compile time, also containing method calls that the partial evaluator cannot inline. As all other TPC-H queries also feature similar properties (most use grouping, sorting, and subselects), we expect similarly poor results for those.

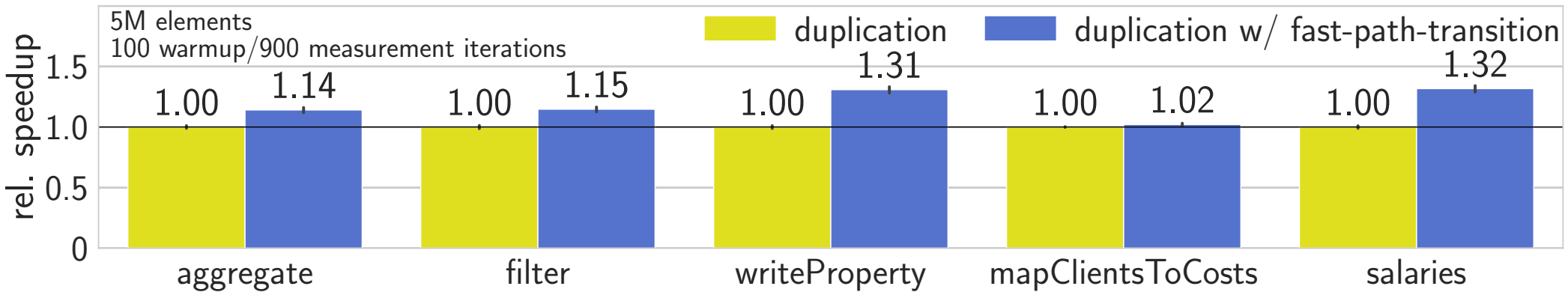

**Figure 11** Results of the slow-to-fast-path-transition relative to GraalVM JavaScript without storage transformation *(higher is better)*

## D Slow-to-Fast-Path Transition

To evaluate our refinement from Section 4.3, we executed some of the microbenchmarks with an array size of 5M. As the performance improvement due to the transition only comes into effect when a method is executed with arrays that are transformed within the method, in each benchmark, we initialize such an array and then execute the query method (e.g., aggregate, filter) *once*. This whole process was repeated 1000 times with the first 100 iterations used as warm-up. We only measured the duration of the query method as the array initialization would otherwise skew the results. Figure 11 shows that most queries indeed show improvements.

## E Storage Transformation Overhead

Especially on smaller workloads, query performance is heavily impacted by the transformation costs induced by our approach. To measure this overhead, we executed some of our microbenchmarks with different numbers of query applications (1, 100, 1000) on arrays of 1M objects. Hence, for each benchmark configuration, the array is transformed into a columnar array exactly once. We then measured the transformation time as well as the total query execution time. Table 5 contains an evaluation of the transformation costs on a number of our microbenchmarks. It shows how especially for simple queries (e.g., *aggregate*) the transformation costs make up a significant part of the overall execution time, which, however, are amortized with an increasing number of query applications.

**Table 5** Transformation overhead

| Query | Iterations (#) | Total (ms) | Query (ms) | Transformation (ms) | (%) |
|---|---|---|---|---|---|
| **aggregate** | 1 | 1300 | 645 | 655 | 50.4 |
| | 100 | 1653 | 991 | 662 | 40.0 |
| | 1000 | 2202 | 1521 | 681 | 30.9 |
| **filter** | 1 | 1717 | 1056 | 661 | 38.5 |
| | 100 | 4630 | 3959 | 671 | 14.5 |
| | 1000 | 22958 | 22296 | 662 | 2.9 |
| **writeProperty** | 1 | 1488 | 907 | 581 | 39.0 |
| | 100 | 1944 | 1362 | 582 | 29.9 |
| | 1000 | 3060 | 2517 | 543 | 17.7 |
| **salaries** | 1 | 2061 | 1554 | 507 | 24.6 |
| | 100 | 3454 | 2925 | 529 | 15.3 |
| | 1000 | 11240 | 10716 | 524 | 4.7 |

## F Octane Benchmark Details

Figure 12 shows how our approach performs on workloads that do not feature suitable arrays. For this, we chose the Octane [16] benchmark suite. While the suite is officially retired, it is still frequently used for compiler evaluation. As the results show, our approach does not cause significant overhead for common workloads in terms of performance, memory usage, and installed code size. The one exception is the *splay* benchmark, where our approach causes a code size increase of around 45 % and a compile time increase of around 35 % but without impact on the performance.

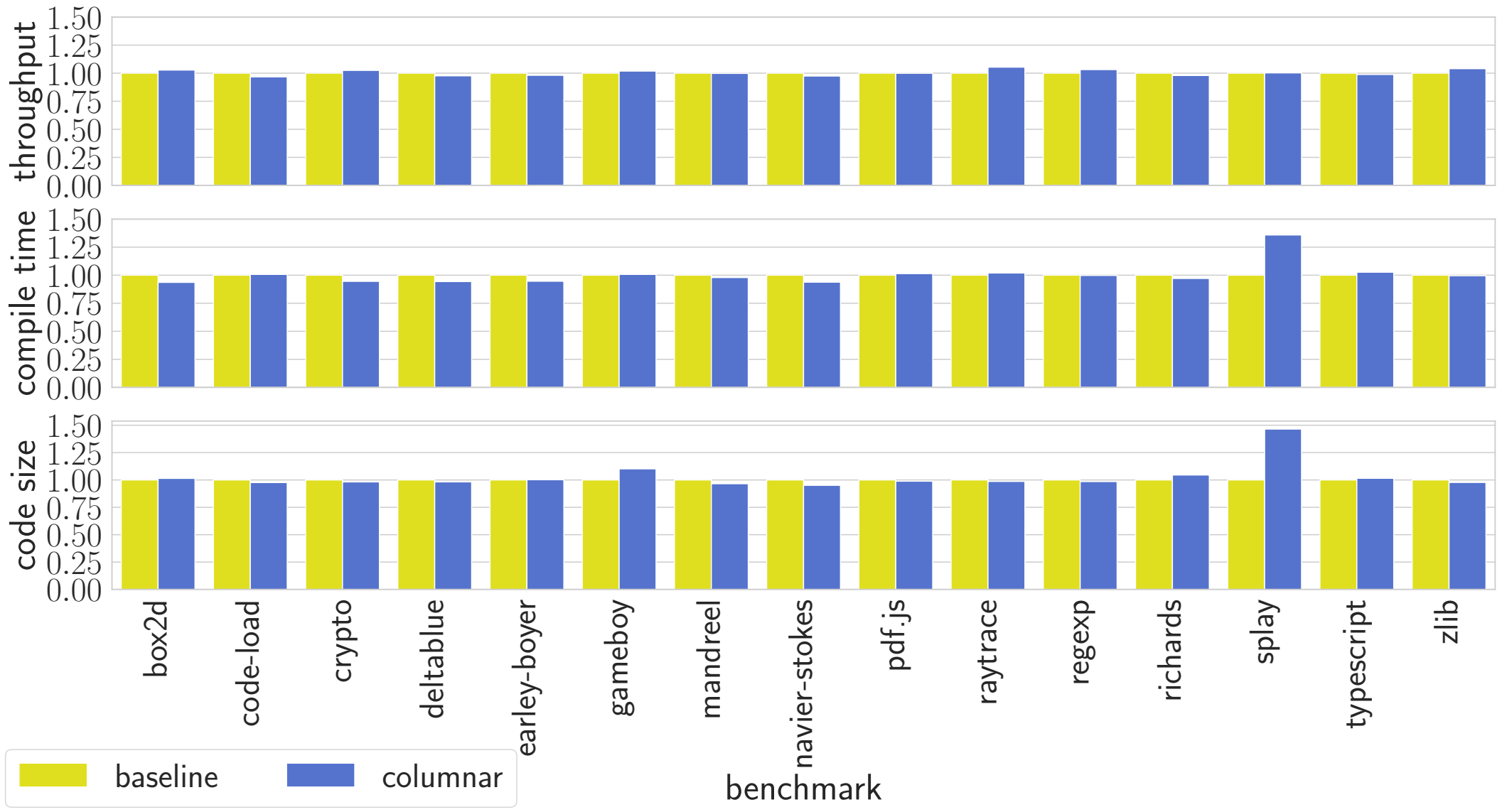

**Figure 12** Octane Benchmark Performance: Throughput (higher is better), compile time (lower is better), code size (lower is better)

## About the authors

**Sebastian Kloibhofer** is a PhD student at the Johannes Kepler University in Linz, Austria. Contact him at sebastian.kloibhofer@jku.at.

**Lukas Makor** is a PhD student at the Johannes Kepler University in Linz, Austria. Contact him at lukas.makor@jku.at.

**David Leopoldseder** is a researcher at Oracle Labs. Contact him at david.leopoldseder@oracle.com

**Daniele Bonetta** is a researcher at Oracle Labs. Contact him at daniele.bonetta@oracle.com

**Lukas Stadler** is a researcher at Oracle Labs. Contact him at lukas.stadler@oracle.com

**Hanspeter Mössenböck** is a professor at the Johannes Kepler University in Linz, Austria. Contact him at hanspeter.moessenboeck@jku.at